\documentclass[12pt, fleqn]{article}

\usepackage{amssymb}
\usepackage{amsmath}
\usepackage{graphicx} 
\usepackage{float}
\usepackage{caption}
\usepackage[margin=1in]{geometry}

\title{A Low Cost Duffing Oscillator}
\author{David Meer, Eric Myers, Richard Halpern }

\begin{document}
\setlength{\parskip}{0.2cm}
\setlength{\parindent}{0pt}
\maketitle
\begin{center}
\textit{Department of Physics and Astronomy, State University of New York at New Paltz}\end{center}
\begin{abstract}
\noindent We describe an undergraduate project to build a Duffing oscillator.  Although the  ultimate goal was to  demonstrate chaos, the overriding consideration was cost, as this was during COVID-19, and budgets were frozen.  Previous designs  used expensive parts that were out of the question in terms of cost.  An inexpensive design whose centerpiece is a video tracking system is presented, along with some comments about aspects of the theory concerning the bi-stable case.
\end{abstract}
\section{Introduction}
In its most general form, the Duffing oscillator is described by the following equation:
\begin{equation}
\ddot{x}+\gamma \dot{x} \pm \alpha x\pm\beta x^3=F\cos(\omega_F t)
\label{duff}\end{equation}
where $\alpha,~\beta,~\gamma,~\omega,~\text{and}~F$ are are taken to be positive. Here $ x=x(t)$ is the position, with the first and second derivatives denoting velocity and acceleration, respectively.  For purposes of this discussion, it is convenient to think of the oscillating object as having unit mass. Then $\ddot x$ is the net force, and by Newton's second law we have:
\[
\ddot{x}= -\gamma \dot{x} \mp \alpha x \mp\beta x^3+F\cos(\omega_F t)
\]
The $\gamma$ term is a friction force that is assumed to depend on the velocity.  The $\alpha$ and $\beta$ terms are  linear and nonlinear restoring forces, respectively, while the term on the right is a  sinusoidal driving term with amplitude  $F$ and angular frequency $\omega_F$. 

\begin{figure}[H]
	\includegraphics[width=.48\linewidth]{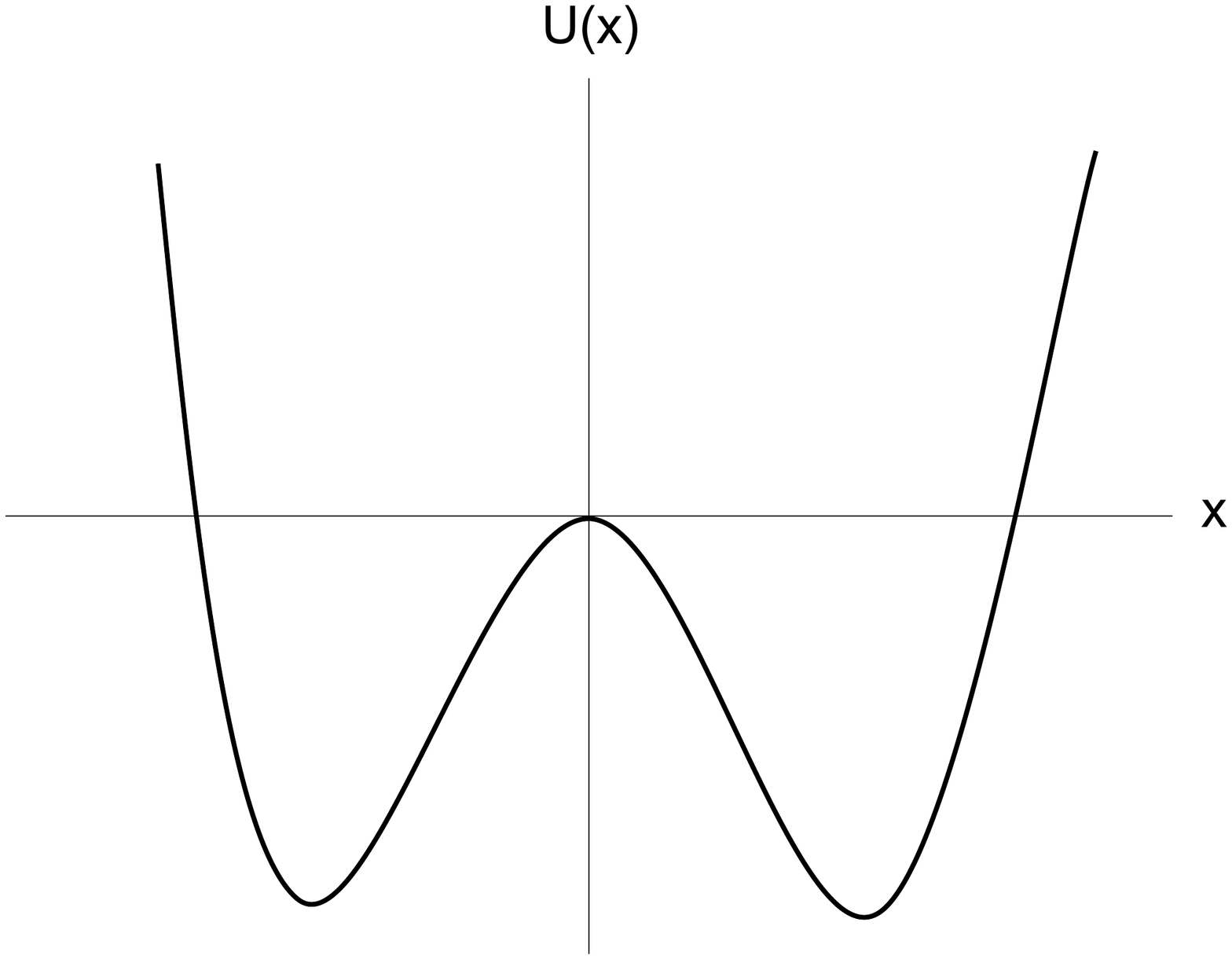}\hfill
	\includegraphics[width=.48\linewidth]{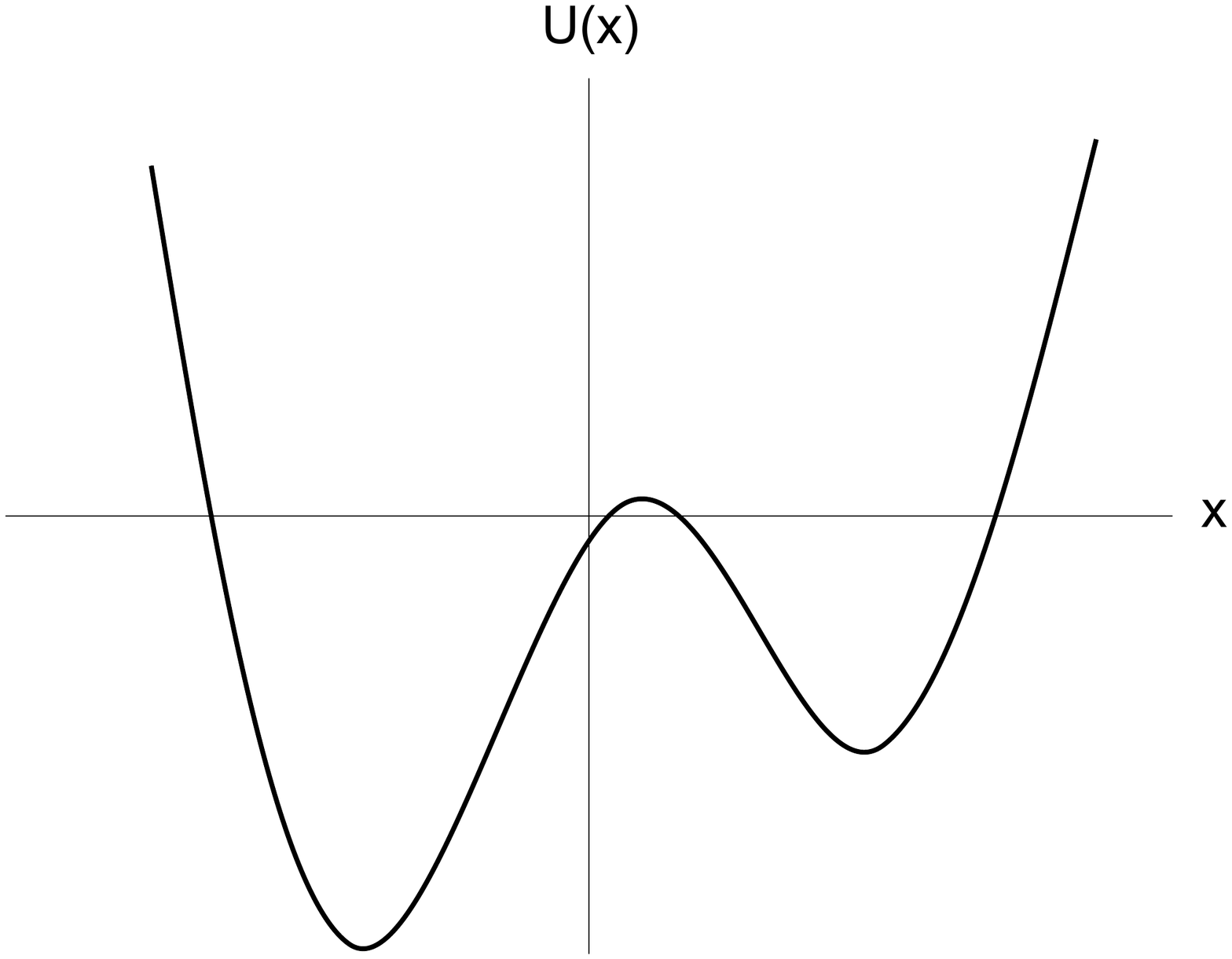}
	\vspace{-0.4cm}
	\caption{Graph of U(x) in equation (\ref{pot}) at two different times.}
	\label{double}
\end{figure}

When the $\alpha$ and $\beta$ terms are both positive, the equation is referred to as the {\em hardening} case.  When the $\alpha$ term is positive and the $\beta$ term is negative, it is called the {\em softening} case.  We are interested in equation (\ref{duff}) where the linear restoring term has the minus sign and the nonlinear restoring term has the plus sign.  This is called the {\em bi-stable} case:
\begin{equation}
\ddot{x}+\gamma \dot{x} - \alpha x + \beta x^3=F\cos(\omega_F t)
\label{bistable}\end{equation}
   For the special case where there is no damping we have:
\begin{equation}
\ddot{x}=  \alpha x - \beta x^3+F\cos(\omega_F t)
\label{eq2}\end{equation}
The right side of equation (\ref{eq2}) is derivable from a potential, $U(x)$:
\begin{equation}\ddot x = -\frac{dU(x)}{dx}~~~~~~~~\text{where}~~U(x) = -\frac{\alpha}{2}x^2+\frac{\beta}{4}x^4-F\cos(\omega_F t)\cdot x\label{pot}\end{equation}
When the driving force is zero, the potential has the form of a double well (figure \ref{double}, left).   The effect of the driving force is to skew the well, an instance of which is shown in figure \ref{double} at the right.
\vspace{-0.3cm}
 The locations of the fixed points at some arbitrary time $\tau$ are given by:    
\[
x_1=2\sqrt[6]{\frac{\alpha^3}{27\beta^3}}\cos\left(\frac{\cos^{-1}\left(\frac{3F\cos(\omega_F \tau)}{2}\sqrt{\frac{3\beta}{\alpha^3}}\right)}{3}\right)\hspace{0.94cm}
\]
\[
x_2=2\sqrt[6]{\frac{\alpha^3}{27\beta^3}}\cos\left(\frac{\cos^{-1}\left(\frac{3F\cos(\omega_F \tau)}{2}\sqrt{\frac{3\beta}{\alpha^3}}\right)+ 2\pi}{3}\right)\]
\[x_3=2\sqrt[6]{\frac{\alpha^3}{27\beta^3}}\cos\left(\frac{\cos^{-1}\left(\frac{3F\cos(\omega_F \tau)}{2}\sqrt{\frac{3\beta}{\alpha^3}}\right)- 2\pi}{3}\right)\]
The locations $x_1$ and $x_2$ are  stable fixed points; $x_3$ is an unstable fixed point.  When the forcing term is set to zero, the above expressions reduce to the values calculated by others [6]:
\[ x_1=\sqrt{\frac{\alpha}{\beta}},~~~~x_2=-\sqrt{\frac{\alpha}{\beta}},~~~~x_3 = 0\]
If the initial energy is positive then the object can jump from one well to the other.  In the absence of friction, this {\em inter-well} motion continues indefinitely.   The effect of the damping force is to dissipate energy so that even if a jump is initially possible,  the object will eventually be trapped in one of the two wells. We refer to motion confined to a single well as {\em intra-well} motion. 

\section{Analysis} 

The  unforced, undamped Duffing equation can be solved analytically using the Jacobi elliptic functions  {\rm cn} and {\rm dn}, depending on the case.  For example, in the bi-stable case we have: 
\begin{subequations}
\begin{align}
 \text{inter-well motion}:~~x_c(t)=A\,\textrm{{\rm cn}}(\omega_c\,t + \theta, M_c)\hspace{0.1cm}\\
  \text{intra-well motion}:~~x_d(t)=A\,\textrm{{\rm dn}}(\omega_d\,t + \theta, M_d)
  \end{align}
  \end{subequations}
Here, $A$ denotes amplitude; it is constant throughout the motion {\em and the same for both the {\rm {\rm cn}} and {\rm {\rm dn}} functions.}  What is the situation when damping is included?  Is there still a single amplitude function for both {\rm cn} and {\rm dn} cases?  We certainly expect the amplitude to decrease with time; previous research into the damped hardening and softening  cases [6], [7] has shown that the amplitude is basically a decaying exponential:
 \begin{equation}A\sim e^{-\frac{\gamma}{2}t}\label{eq3}\end{equation}
 
 \begin{figure}[H]
	 \includegraphics[width=.48\linewidth]{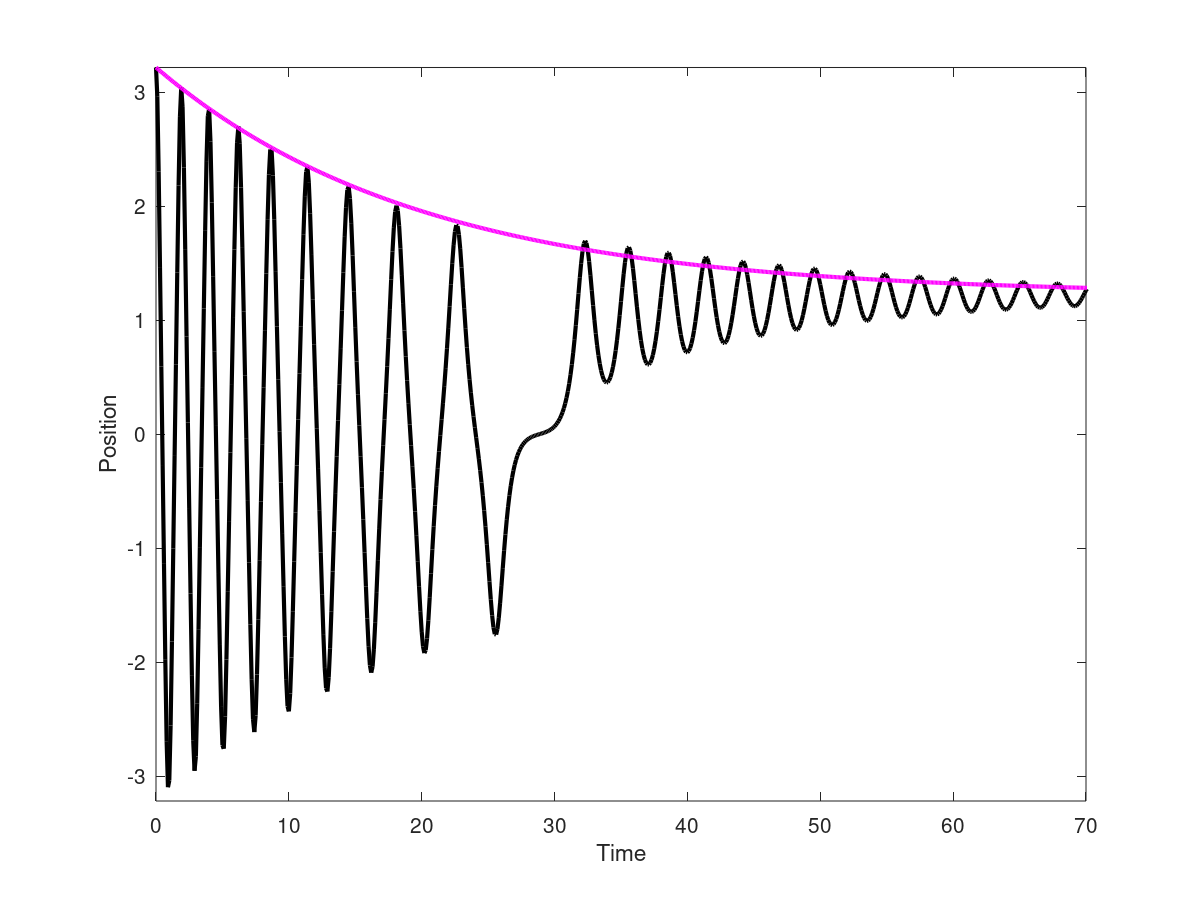}\hfill
	 \includegraphics[width=.48\linewidth]{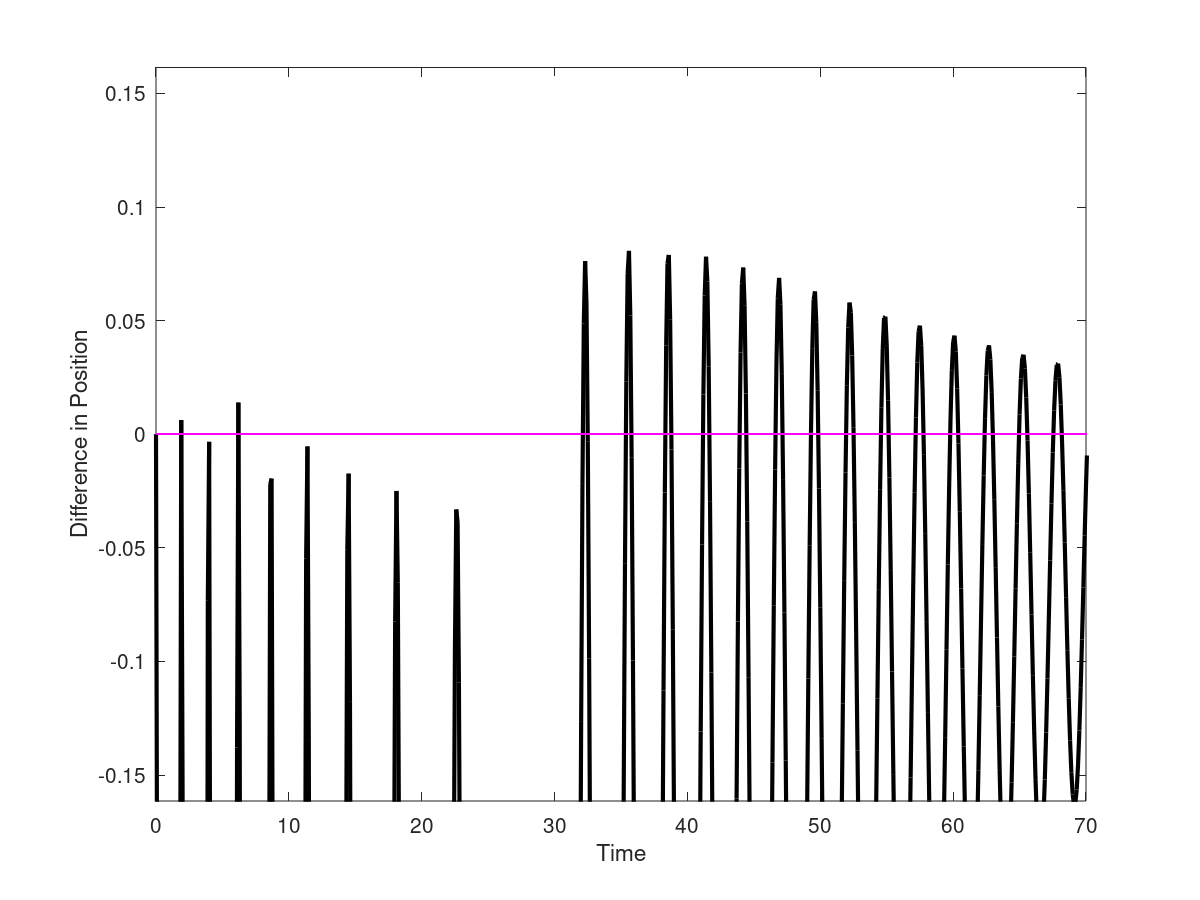}
	 \caption{Left: Comparison of trajectory peaks (black) with the amplitude function (magenta). Right: Difference between trajectory peaks and the amplitude function.  The horizontal line represents zero difference.}
	 \label{f6}
 \end{figure}
 
 This can't be the case for the bi-stable oscillator since the amplitude must end up at  one of the double well fixed points, not zero.  We speculated that an amplitude function analogous to (\ref{eq3}) for the bi-stable case might look like this:
 \[A(t)=(A_0-x^*)e^{-\frac{\gamma}{2}ht}+x^*\]
  Here, $A_0$ is the initial amplitude,  $x^*$ is the position of  the bottom of the well, and  $h$ is a parameter that we can adjust to get the best fit for a given set of initial conditions.  Even with the optimum $h$ value  there is a problem with the result.  In figure \ref{f6} (left) we see plots of amplitude $vs$ time (magenta), and  position $vs$ time for the pendulum (black).  We would expect the amplitude graph to just touch each peak over the entire interval of the motion.   That is not the case, as figure \ref{f6} (right) demonstrates.  Here, the {\em difference} between the amplitude value and each trajectory peak is  plotted.  Ideally, each peak should be at the horizontal line.  The inter-well peaks are  below where one would expect them to be, while the intra-well peaks are  above where one would expect.  This strongly suggests that a single exponential function does not capture the actual behavior of the amplitude.  It is questionable that  combinations of exponentials in the two different regimes will do the trick. This is an issue that certainly needs further investigation.

Another problem we ran into was a discrepancy between our expression for the amplitude of the hardening oscillator,  equation (\ref{ours})
  \begin{equation}A^2 = -\frac{\alpha}{\beta} + \frac{(\alpha+\beta\,x_0^2)}{\beta}\sqrt{ 1 +\frac{2\beta\,v_0^2}{(\alpha+\beta\,x_0^2)^2}}\label{ours}\end{equation}
  and the expression given in [6], equation (\ref{hers}):  
\begin{equation}A^2 = -\frac{\alpha}{\beta} + \frac{(\alpha+\beta\,x_0^2)}{\beta}\sqrt{ 1 +\frac{2\beta\,v_0^2}{\alpha}}\label{hers}\hspace{1.1cm}\end{equation}
We present the  derivation of our result in equation (\ref{ours}) in an Appendix.

\section{Construction}

Previous implementations of the Duffing oscillator suitable for an undergraduate project [4], [5] used expensive tools for measuring the motion of the system.  Our objective, which was motivated by a strict COVID-19 budget freeze, was to construct a device as inexpensively as possible.   That meant building as much as possible from scratch, using materials that we had on hand, and figuring out an economical way to track the motion.

\begin{figure}[H]
	\begin{center}
	 \includegraphics[width=11cm, height=7.5cm]{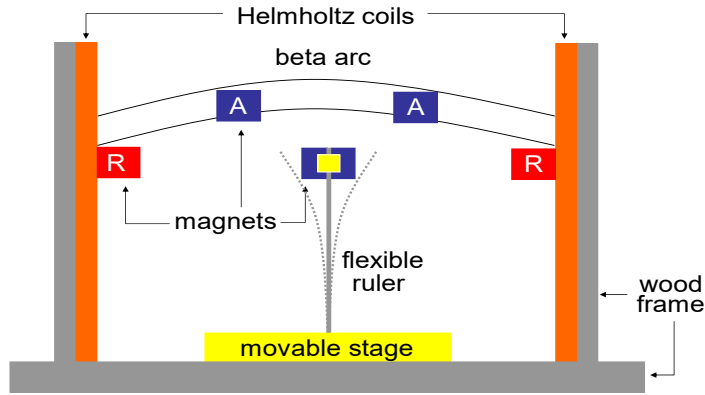}
	  \vspace{-0.8cm}
	 \caption{Schematic diagram of the oscillator.}
	 \vspace{-0.8cm}
	\label{schem}
	\end{center}
 \end{figure}

A schematic representation of the apparatus is shown in figure \ref{schem}.   The ``pendulum'' is a vertically mounted flexible plastic ruler with a pair of permanent magnets at the top.  (The attractive force of the two magnets is strong enough so that they hold firmly to the ruler.)   The bottom of the ruler is fastened to a movable wooden stage that allows for small lateral adjustments in its position.  The dotted lines indicate how the ruler can flex.  There are four other pairs of permanent magnets, all at fixed locations. Each pair is held in place the same way as described for the pendulum.  

\vspace{-0.4cm}
 Two of them, which attract the pendulum and are labeled ``A'', are mounted on a 3D-printed strut in the shape of an arc that roughly parallels the arc traced out by the pendulum.   We call this a ``beta arc'' because we can change the value of $\beta$  by changing the position of these magnets on the strut.  Their placement  also determines the locations of the two potential minima seen in figure \ref{double}. 
 The two magnets labeled ``R'', located at the base of the beta arc,  repel the magnets on the pendulum; they enhance the restoring force when pendulum swings far enough to pass beyond either of the potential minima.  
 
 The small yellow rectangle on the pendulum top is a piece of yellow tape which is used for tracking the motion, as described in the next section.
 
The aforementioned components are mounted between a pair of Helmholtz coils that were constructed from scratch.  They were wound on a pair of 26 inch bicycle wheels that had been discarded by a local bike shop.  A bicycle wheel is a convenient frame on which to wind a coil.  It is rigid, easily mounted, and has a deep channel in which many turns of wire can be held.  The coils are connected to a function generator whose (sinusoidal) output is increased by a small power amplifier.  The  resulting sinusoidal magnetic field between the coils provides the variable driving force for the oscillator. 

\begin{figure}[H]
	\begin{center}
	 \includegraphics[width=14cm, height=9cm]{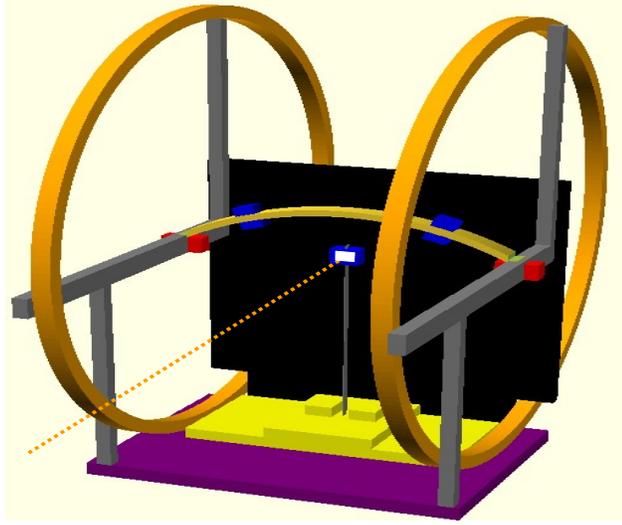}
	 \caption{3D view of the apparatus.}
	 \label{scad}
	\end{center}
 \end{figure}

Figure \ref{scad} shows a 3D view of the apparatus as rendered using OpenSCAD software.  There is no physical significance to the color scheme; it is simply for ease of viewing.  

 When we first planned the construction of the Helmholtz coils, the bike wheels seemed attractive  because they gave us enough space between them to set up the video system.  The fact that they were free also di{\rm dn}'t hurt.  Budgetary considerations forced us to use the only magnet wire we had available, which was thin and unmarked.   We used 123 turns of this wire giving a total resistance of 60 $\Omega$.  We were able to drive a 45 mA current through the coils, and while the magnetic field at the center was quite uniform, it was only 0.2 mT, too small to  drive the pendulum effectively.  Hence, the only choice we had to test our tracking system was to do it on the unforced pendulum, the details of which are given in the next section.
\vspace{-0.5cm}

\section{Video Tracking}
\begin{figure}[H]
	\includegraphics[width=.40\linewidth]{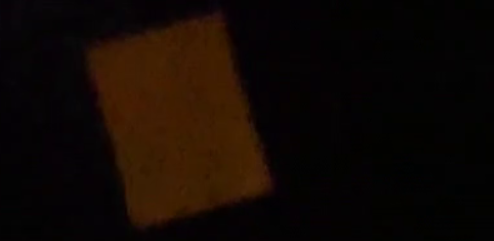}\hfill
	\includegraphics[width=.40\linewidth]{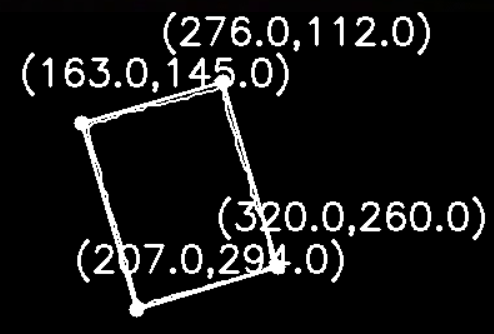}
	\caption{Left: Yellow tape that is tracked by the video system.  The image is very faint.  Right: The output of the video software for the image at the left.  The numbers are the coordinates of the corners of the isolume.}
	\label{video}
\end{figure}

Data are collected using a novel video tracking scheme.  A bright square of tape is affixed to the center of the pendulum head.   The apparatus is situated in a dim room, and a  cell phone set to record video at 240 FPS is mounted on a tripod and focused on the tape.  The dotted line in figure \ref{scad} is the line of sight for the video camera.   At a high frame rate, the only significant light in the video comes from the light reflected off the tape (figure \ref{video}, left). The video is then cropped to block any light not coming from the tape. The result is converted to grayscale, and isolumes are plotted. The isolume encapsulating the largest area is selected, and a box is drawn around this area (figure \ref{video}, right). The center of this box is recorded, and an angle is computed from the bottom of the data frame. We use this angle to track the relative motion of the oscillator.
 
\begin{figure}[H]
	\includegraphics[width=.48\linewidth]{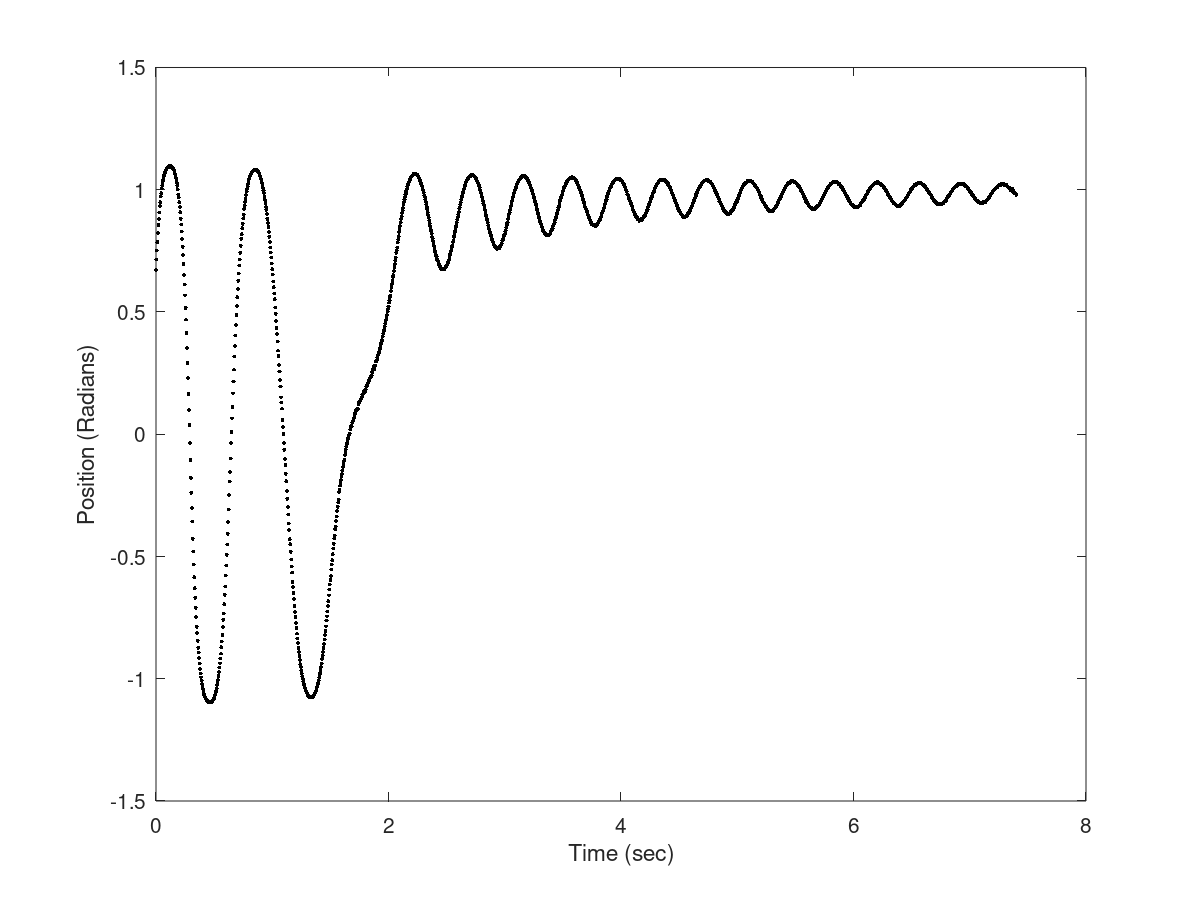}\hfill
	\includegraphics[width=.48\linewidth]{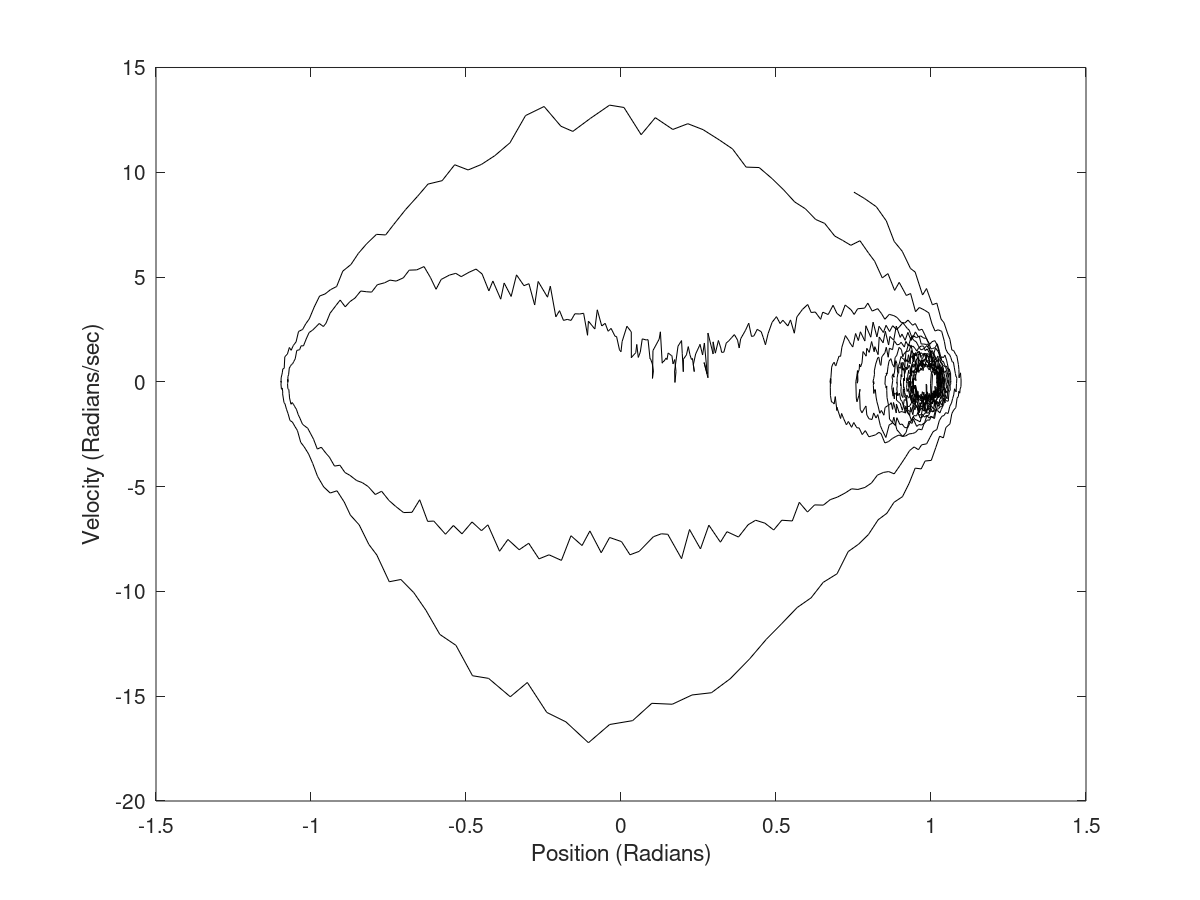}
	\caption{Left: A plot of the trajectory of the oscillating pendulum.  Right: The phase plot of the same trajectory.}
	\label{f5}
\end{figure}

We tested out the video tracking system starting with the unforced oscillator.  Some results are shown in figure \ref{f5}.  At the left is a position vs time plot, and at the right is the corresponding phase portrait.   The initial amplitude is large enough so that the pendulum can  jump from one well to the other, but very soon it loses enough energy so that it gets trapped in the right hand part of the double well, eventually tailing off to zero.  One can see a slowing down at about t = 2 s, beyond which the pendulum no longer has the energy to jump from one well to the other.   In both of these results, the behaviors were what we expected, and provided evidence that the video tracking system was working properly.

\section{Future Work}
We plan to continue theoretical work on two issues:
\begin{itemize}
\item We want to find a function that predicts how much initial energy the damped, unforced pendulum can have yet be incapable of inter-well oscillation.
\item We know the trajectory as a function of amplitude, but we don't know how the amplitude changes with time.  Hence, we would like to find a function that gives us the amplitude as a function of time.
\end{itemize}
Also, we now know that we can use the video system with smaller diameter coils; with thicker wire and the smaller coils we should be able to produce a stronger Helmholtz field.  That will open up the possibility of using this system to investigate chaotic motions.
 
Despite the cost constraints  that we noted, we made toward our goal of building an inexpensive Duffing oscillator suitable for classroom demonstration and undergraduate research.

\section{Acknowledgments}
We would like to thank Dr. Catherine Herne and Joe Spiconardi for their help in this project. Dr. Herne spearheaded the data collection method and created an initial version, and Joe Spiconardi helped in the initial design and construction.  We are also grateful for the general support of the  Department of Physics and Astronomy at SUNY New Paltz.

\section{Appendix}

The hardening form of the unforced, undriven Duffing oscillator and its solution are
\[\text{equation}:~\ddot x + \alpha\,x+\beta\,x^3 = 0~~~~~~~\text{solution}:~x(t) = A\,{\rm cn}(u,m)\]
Here, $u = \omega\, t + \theta$ and $m$ is the elliptic modulus.  For this case, it is known that
\[m\omega^2 = \frac{1}{2}\beta\,A^2~~~~~\text{and}~~~~~\omega^2 = \alpha + \beta\,A^2\]
In this derivation, we will make use of two identities:
\[{\rm sn}^2(u,m) + {\rm cn}^2(u,m) = 1~~~~~~{\rm dn}^2(u,m) = 1 - m\,{\rm sn}^2(u,m)\]
We will also need the time derivative of {\rm cn}:
\[\frac{d[{\rm cn}(u,m)]}{dt}=\frac{d[{\rm cn}(u,m)]}{du}\cdot \frac{du}{dt}=-{\rm sn}(u,m)\,{\rm dn}(u,m)\cdot \omega\]

Start with the expression for the velocity:
\[v(t) = \dot x(t) = A\frac{d[{\rm cn}(u,m)]}{dt} = -A\,\omega\,{\rm sn}(\omega\, t + \theta,m)\,{\rm dn}(\omega\, t + \theta,m)\]
Let the initial velocity be denoted by $v_0$, and the initial position be denoted by $x_0$.  Then
\[x(0) = x_0=A\,{\rm cn}(\theta,m)~~~~~v(0) = v_0 = -A\,\omega\,{\rm sn}(\theta,m)\,{\rm dn}(\theta,m)\]For brevity and clarity in the notation, we will henceforth write {\rm sn}($\theta,m)$ as simply {\rm sn}, with the same idea holding for {\rm cn} and {\rm dn}. Then
\[  v_0 = -A\,\omega\,{\rm sn}\,{\rm dn} ~~~~~~~~~\text{also:}~~{\rm cn} = \frac{x_0}{A}  \]
Now square the initial velocity and use the identities to make substitutions:
\[ v_0^2 =  \omega^2\, A^2\, {\rm sn}^2\, {\rm dn}^2=\omega^2\, A^2\, [1-{\rm cn}^2]\, [1-m\, (1-{\rm cn}^2)] \]
\[\hspace{0.5cm} =\omega^2\, A^2\, [1-{\rm cn}^2]\, [1- m +m\, {\rm cn}^2] \]
\[\hspace{0.5cm}=\omega^2\, A^2\, - m\omega^2\, A^2\,  +m\omega^2\, A^2\, {\rm cn}^2- \omega^2\, A^2\, {\rm cn}^2+m\, \omega^2\, A^2\, {\rm cn}^2-m\,\omega^2\, A^2\,  {\rm cn}^4\]
 For the ${\rm cn}$ solutions, $m\omega^2 = \frac{1}{2}\,\beta \,A^2$, so substituting gives
 \[ v_0^2 =\omega^2\, A^2\, - \frac{1}{2}\,\beta \,A^4 +\frac{1}{2}\,\beta \,A^4\, {\rm cn}^2- \omega^2\, A^2\, {\rm cn}^2+\frac{1}{2}\,\beta \,A^4{\rm cn}^2-\frac{1}{2}\,\beta \,A^4\,  {\rm cn}^4\]
Substitute for ${\rm cn}$:
 \[v_0^2=\omega^2\, A^2\, - \frac{1}{2}\,\beta \,A^4 +\frac{1}{2}\,\beta \,A^4\, \left(\frac{x_0}{A}\right)^2- \omega^2\, A^2\, \left(\frac{x_0}{A}\right)^2+\frac{1}{2}\,\beta \,A^4\left(\frac{x_0}{A}\right)^2-\frac{1}{2}\,\beta \,A^4\,  \left(\frac{x_0}{A}\right)^4\]
 Canceling and combining terms gives:
  \[v_0^2=\omega^2\, A^2\, - \frac{1}{2}\,\beta \,A^4 +\beta \,A^2\, x_0^2- \omega^2\, x_0^2-\frac{1}{2}\,\beta \,x_0^4\]
  We also know that $\omega^2 = \alpha + \beta\,A^2$, so
  \[v_0^2=(\alpha + \beta\,A^2)\, A^2\, - \frac{1}{2}\,\beta \,A^4 +\beta \,A^2\, x_0^2- (\alpha + \beta\,A^2)\, x_0^2-\frac{1}{2}\,\beta \,x_0^4\]
  Expand
    \[v_0^2=\alpha \, A^2+ \beta\,A^4\, - \frac{1}{2}\,\beta \,A^4 - \alpha\, x_0^2 -\frac{1}{2}\,\beta \,x_0^4=\alpha \, A^2+  \frac{1}{2}\,\beta \,A^4 - \alpha\, x_0^2 -\frac{1}{2}\,\beta \,x_0^4\]
   Arrange in descending powers of ``A", then  multiply by $2/\beta$:
    \[ 0 =  \frac{1}{2}\,\beta \,A^4+\alpha \, A^2 - \left(\alpha\, x_0^2 +\frac{1}{2}\,\beta \,x_0^4 + v_0^2\right)=A^4+\frac{2\alpha }{\beta}\, A^2 - \left(\frac{2\alpha\, x_0^2}{\beta} +x_0^4 + \frac{2v_0^2}{\beta}\right)\]
   More algebra:
      \[A^4+\frac{2\alpha}{\beta}\, A^2 - \left(\frac{2\alpha\beta\, x_0^2}{\beta^2} +\frac{\beta^2\,x_0^4}{\beta^2} + \frac{2\beta\,v_0^2}{\beta^2}\right)=A^4+\frac{2\alpha}{\beta}\, A^2 - \frac{2\alpha\beta\, x_0^2+\beta^2\,x_0^4 + 2\beta\,v_0^2}{\beta^2}=0\]
      Now solve the quadratic equation in $A^2$:
      \[A^2 = \left[-\frac{2\alpha}{\beta} +\sqrt{ \frac{4\alpha^2}{\beta^2}+4\,\frac{2\alpha\beta\, x_0^2+\beta^2\,x_0^4 + 2\beta\,v_0^2}{\beta^2}}\right] \div 2\]
      Yes, more algebra:
       \[A^2 = -\frac{\alpha}{\beta} + \frac{1}{\beta}\sqrt{ \alpha^2+\,2\alpha\beta\, x_0^2+\beta^2\,x_0^4 + 2\beta\,v_0^2}= -\frac{\alpha}{\beta} + \frac{1}{\beta}\sqrt{ (\alpha+\beta\,x_0^2)^2 + 2\beta\,v_0^2}\]
       More rearranging, and the final result:
        \[A^2 = -\frac{\alpha}{\beta} + \frac{(\alpha+\beta\,x_0^2)}{\beta}\sqrt{ 1 +\frac{2\beta\,v_0^2}{(\alpha+\beta\,x_0^2)^2}}\]
\section{Citations}
\begin{enumerate}
  \item T.J. Kazmierski, S. Beeby, {\em Energy Harvesting Systems: Principles, Modeling, and Applications}, Springer, New York, 2011
  \item M.I. Dykman, A.L. Velikovich, G.P. Golubev, D.G. Luchinski, S.V. Tsuprikov, {\em Stochastic Resonance in an all-optical passive bistable system}, Sov. Phys. JETP Lett. 53 (1991) 193-197
  \item L.N. Virgin, {\em Vibration of Axially Loaded Structures}, Cambridge University Press, Cambridge, 2007
  \item Krantz, Richard, et al. {\em Nonlinear Dynamics on the Cheap in the Junior Laboratory}, 2015, Denver, Department of Physics, Metropolitan State University of Denver.
  \item Berger, J. E. and Nunes, G. (1997). {\em A mechanical Duffing oscillator for the undergraduate laboratory},  American Journal of Physics, 65, 841–846. 
  \item Kovacic, Ivana, and Michael J. Brennan. {\em The Duffing Equation: Nonlinear Oscillators and Their Phenomena}, Wiley, 2011.
  \item Johannessen, Kim. (2015). {\em The Duffing oscillator with damping}, European Journal of Physics. 36. 10.1088/0143-0807/36/6/065020.
  \item Chunlin Zhang, R.L. Harne, Bing Li, K.W. Wang, {\em Reconstructing the transient, dissipative dynamics of a bistable Duffing oscillator with an enhanced averaging method and Jacobian elliptic functions}, ,International Journal of Non-Linear Mechanics,Volume 79,2016,Pages 26-37,ISSN 0020-7462
\end{enumerate}

\end{document}